\documentclass[letterpaper,twocolumn,english,nofootinbib]{revtex4}
\usepackage{ae}
\usepackage{aecompl}
\usepackage[T1]{fontenc}
\usepackage[latin1]{inputenc}
\usepackage{amsmath}
\usepackage{amssymb}

\makeatletter
\usepackage{babel}
\makeatother
\begin{document}

\title{Observability of the total inflationary expansion }

\author{Sergei Winitzki}

\affiliation{Arnold Sommerfeld Center, Department of Physics, Ludwig-Maximilians
University, Munich, Germany}

\date{\today}

\begin{abstract}
I consider the question of possible observability of the total number
of $e$-folds accumulated during the epoch of inflation. The total
number of observable $e$-folds has been previously constrained by
the de Sitter entropy after inflation, assuming that the null energy
condition (NEC) holds. The NEC is violated by upward fluctuations
of the local Hubble rate $H$, which occur with high probability in
the fluctuation-dominated regime of inflation. These fluctuations
lead at late times to the formation of black holes and thus limit
the observability of inflationary evolution. I compute the average
number $\left\langle \Delta N\right\rangle $ of $e$-folds accumulated
during the last NEC-preserving fragment of the inflationary trajectory
before reheating. This is the maximum number of inflationary $e$-folds
that can be observed in principle through measurements of the CMB
at arbitrarily late times (if the dark energy disappears). The calculation
also provides a reasonably precise definition of the boundary of the
fluctuation-dominated regime, with an uncertainty of a few percent.
In simple models of single-field inflation compatible with current
CMB observations, I find $\left\langle \Delta N\right\rangle $ of
order $10^{5}$. This upper bound on the observable $e$-folds, although
model-dependent, is much smaller than the de Sitter entropy after
inflation. The method of calculation can be used in other models of
single-field inflation.
\end{abstract}
\maketitle

\section{Introduction and summary}

Inflation produces primordial metric fluctuations that may be observed
indirectly through CMB measurements such as WMAP~\cite{Bennett:2003bz}.
An observation of CMB at present corresponds to the measurement of
the inflaton evolution about 60 $e$-folds before reheating~\cite{Liddle:2003as}.
Assuming that CMB measurements will be possible at indefinitely late
times, one might hope to deduce information about arbitrarily early
stages of inflation. Of course, late-time acceleration (persistent
{}``dark energy'') can make it impossible to observe CMB at very
late times~\cite{Krauss:2007nt}. There is, however, another limit
on our ability to see towards the past. This limit is caused by violations
of the null energy condition (NEC) during inflation.

To make the following arguments more specific, let us consider a model
of inflation driven by a canonical, minimally coupled scalar field
$\phi$ such that the field evolves from a large initial value $\phi_{\text{in}}$
(perhaps near the Planck boundary $\phi_{\text{Pl}}$) to the reheating
point $\phi=\phi_{*}$. A typical model of this type has the inflaton
action\begin{equation}
\int d^{4}x\sqrt{-g}\left(\frac{1}{2}\phi_{,\mu}\phi^{,\mu}-V(\phi)\right).\label{eq:inflaton action}\end{equation}
 We assume that the inflaton potential $V(\phi)$ grows monotonically
with $\phi$ and that the slow-roll approximation is valid. In models
of chaotic type, e.g.~$V(\phi)\propto\phi^{2n}$, we then expect
that $\phi_{*}\ll\phi_{\text{in}}\lesssim\phi_{\text{Pl}}$, and that
$\phi_{\text{in}}$ is deep in the fluctuation-dominated regime. 

The evolution of $\phi$ during inflation can be pictured as a random
walk superimposed on a deterministic drift towards $\phi=\phi_{*}$
\cite{Vilenkin:1983xq,Starobinsky:1986fx,Linde:1986fd}. The random
walk can be modeled as {}``diffusion'' in $\phi$ space with the
diffusion coefficient \begin{equation}
D(\phi)\equiv\frac{H^{3}}{8\pi^{2}},\quad H(\phi)\equiv\frac{8\pi G}{3}V(\phi)\equiv\frac{8\pi}{3M_{\text{Pl}}^{2}}V(\phi),\end{equation}
while the mean drift velocity is the time derivative $\dot{\phi}$
of the slow-roll evolution,\begin{equation}
\dot{\phi}=v(\phi)\equiv-\frac{H^{\prime}}{4\pi G}=-\frac{H^{\prime}}{4\pi}M_{\text{Pl}}^{2}.\label{eq:phi dot sr}\end{equation}
During the last stages of inflation before reheating, the trajectory
$\phi(t)$ is monotonic ($\dot{\phi}<0$) with nearly unit probability,
although there is always a small probability of an upward fluctuation
($\dot{\phi}>0$). On the other hand, in the fluctuation-dominated
regime an upward fluctuation of $\phi$ has around 50\% probability
because random fluctuations dominate over the slow-roll motion. 

We now note that an upward fluctuation of $\phi$ corresponds to an
upward fluctuation of the local Hubble rate $H$ and thus to a local
violation of the null energy condition (NEC) \cite{Winitzki:2001fc,Vachaspati:2003de}.
A violation of the NEC due to an upward fluctuation of $H$ on a distance
scale $L\sim H^{-1}$ leads to the formation of a Hubble volume that
looks like a black hole from the outside~\cite{Blau:1986cw}, and
to a real black hole when the overdensity enters the local Hubble
horizon at late times after inflation~\cite{Linde:1988zp,Bousso:2006ge}.
This can be understood qualitatively by noting that a Hubble-size
region of approximately de Sitter spacetime with local Hubble parameter
$H$ has exactly the energy density $\Lambda=\frac{3}{8\pi}M_{\text{Pl}}H^{2}$
that corresponds to a black hole with the Schwarzschild radius $H^{-1}$.
An upward fluctuation of $H$ therefore leads to an increase of the
energy density beyond the Schwarzschild limit. In this way, an NEC
violation in the far inflationary past will limit the lifetime of
any future observers who may be trying to perform CMB observations
at very late times. 

The main focus of this paper is an investigation of this limiting
effect of NEC violations. For the sake of this consideration, I will
assume that dark energy eventually decays, so that the late-time universe
is not expanding with acceleration and the local Hubble radius grows
without limit, permitting (in principle) observations of the primordial
density fluctuations on arbitrarily large scales.

It is interesting to determine the time range within which the NEC
can be violated during inflation. For each random inflationary trajectory
$\phi(t)$ there exists a well-defined time of the \emph{last} NEC
violation before reheating, i.e.~a time $t_{\text{N}}$ such that
the NEC is violated around $t=t_{\text{N}}$ but then is not violated
any more. The evolution of $\phi(t)$ before $t=t_{\text{N}}$ is
thus not observable even in principle. On the other hand, the evolution
of $\phi$ after $t=t_{\text{N}}$ is in principle observable: If
primordial fluctuations on a distance scale $L$ are produced at $t>t_{\text{N}}$,
these fluctuations will be observed through the CMB fluctuations at
sufficiently late times when the scale $L$ reenters the Hubble horizon
(we are assuming that the dark energy does not prevent such observations).
Therefore, it is only the statistics of the last NEC-preserving%
\footnote{I talk about {}``NEC-preserving'' rather than about {}``monotonically
decreasing'' trajectories $\phi(t)$ because in models with several
fields ($\phi_{1},...,\phi_{n}$), an NEC violation does not necessarily
entail an upward fluctuation of a particular field $\phi_{k}(t)$. %
} segment of the inflaton trajectory $\phi(t)$ that is --- even in
principle --- accessible to observations.

In the following sections we will compute (within an adequate approximation)
the mean duration of the last NEC-preserving portion of the trajectory
$\phi(t)$ until reheating. When fluctuations are negligible, the
field evolves according to the slow-roll equation~(\ref{eq:phi dot sr}),
and so the time needed for evolving from $\phi=\phi_{1}$ to reheating
(taking into account that $\phi_{*}<\phi_{1}$) is\begin{equation}
\Delta t(\phi_{1},\phi_{*})=\int_{\phi_{*}}^{\phi_{1}}\frac{d\phi}{-v(\phi)}.\end{equation}
(We write $-v$ because the value of $\phi$ decreases with time,
so $v(\phi)<0$.) This formula, however, cannot be used directly to
compute the duration of the last NEC-preserving portion of the trajectory,
for two reasons: First, the NEC-preserving portion of the trajectory
depends on chance and is not confined within a fixed interval, say
$\left[\phi_{*},\phi_{1}\right]$. Second, the initial stages of the
trajectory $\phi(t)$ belong to the fluctuation-dominated regime where
the evolution $\phi(t)$ is not well described by the deterministic
slow-roll equation $\dot{\phi}=v(\phi)$. 

Below we will compute the \emph{average} duration $\left\langle \Delta t_{\text{NEC}}\right\rangle $
of the last NEC-preserving portion of the trajectory $\phi(t)$, where
the average is performed over the ensemble of all comoving trajectories.%
\footnote{Thus we compute the {}``comoving'' average rather than a {}``volume-weighted''
average, which would require more complicated calculations left for
future work. See, e.g., Ref.~\cite{Winitzki:2006rn} for a review
of comoving and volume-weighted averaging prescriptions. %
} Heuristically, we may attempt to determine a value $\phi=\phi_{q}$
such that the duration of the slow-roll trajectory between $\phi=\phi_{q}$
and $\phi=\phi_{*}$ is precisely equal to $\left\langle \Delta t_{\text{NEC}}\right\rangle $,
i.e.~we first compute $\left\langle \Delta t_{\text{NEC}}\right\rangle $
and then \emph{define} $\phi_{q}$ such that\begin{equation}
\Delta t(\phi_{q},\phi_{*})=\int_{\phi_{*}}^{\phi_{q}}\frac{d\phi}{-v(\phi)}=\left\langle \Delta t_{\text{NEC}}\right\rangle .\end{equation}
 This value $\phi_{q}$ can then be interpreted as the boundary between
the fluctuation-dominated and the fluctuation-free regimes. One of
the main results of this paper is a method of computing $\left\langle \Delta t_{\text{NEC}}\right\rangle $;
thus, the boundary $\phi_{q}$ between the fluctuation-dominated and
the fluctuation-free regimes becomes a well-defined quantity. A precise
definition of this boundary is relevant, e.g., for certain measure
prescriptions for regulating eternal inflation~\cite{Linde:2007nm,Linde:2008xf}
as well as for attempts to count the observable degrees of freedom
after inflation~\cite{ArkaniHamed:2007ky,Linde:2009ah}.

The boundary between the fluctuation-dominated and the fluctuation-free
regimes can be characterized in terms of the dimensionless ratio of
$-v\delta t$ (the change of the field $\phi$ due to slow roll during
one Hubble timestep $\delta t\equiv H^{-1}$) to $\sqrt{2D\delta t}$
(the typical fluctuation during the same time),\begin{equation}
b(\phi)\equiv\frac{-v\delta t}{\sqrt{2D\delta t}}=\frac{-2\pi\dot{\phi}}{H^{2}}=\sqrt{\frac{3}{8}\varepsilon_{1}\frac{M_{\text{Pl}}^{4}}{V(\phi)}},\label{eq:b def 0}\end{equation}
where \begin{equation}
\varepsilon_{1}\equiv\frac{M_{\text{Pl}}^{2}}{16\pi}\frac{V^{\prime2}}{V^{2}}\label{eq:epsilon1 def}\end{equation}
is the first slow-roll parameter. We note that $b^{2}(\phi)$ coincides,
for inflationary models of the type~(\ref{eq:inflaton action}),
with the inverse magnitude of the power spectrum of the primordial
scalar fluctuation mode that crossed the Hubble scale at that time,\begin{equation}
P_{S}\approx\frac{1}{4\pi^{2}}\left(\frac{H^{2}}{\dot{\phi}}\right)^{2}=\frac{8}{3\varepsilon_{1}}\frac{V(\phi)}{M_{\text{Pl}}^{4}}=\frac{1}{b^{2}(\phi)}.\label{eq:PS def}\end{equation}
(In the equation above, we neglected the slow-roll corrections since
our final result will only depend logarithmically on $P_{S}$.) The
fluctuation-free regime is characterized by $b(\phi)\gg1$ and the
fluctuation-dominated regime by $b(\phi)\lesssim1$ (i.e.~by primordial
fluctuations of order 1 or larger). However, this qualitative characterization
cannot provide a sharply defined boundary value $\phi_{q}$ separating
the two regimes.

We also note that the order parameter $\Omega$, which was used in
Ref.~\cite{Creminelli:2008es} to characterize the transition between
the presence and the absence of eternal inflation, is related to $b$
by\begin{equation}
\Omega\equiv\frac{2\pi^{2}}{3}\frac{\dot{\phi}^{2}}{H^{4}}=\frac{\pi}{3}b^{2}.\end{equation}
The fluctuation-dominated regime was characterized by the condition
$\Omega<1$ in Ref.~\cite{Creminelli:2008es}, which again corresponds
qualitatively to $b\lesssim1$. 

It will be shown below that the rigorously computed value $\left\langle \Delta t_{\text{NEC}}\right\rangle $
can be approximated as \begin{equation}
\left\langle \Delta t_{\text{NEC}}\right\rangle \approx\int_{\phi_{*}}^{\phi_{\text{Pl}}}\frac{d\phi}{-v(\phi)}f(0;\phi),\end{equation}
where $f(0;\phi)$ is the probability of the event that an inflationary
trajectory $\phi(t)$ starting at $t=0$ with the given value of $\phi$
will \emph{never} violate the NEC. It will be found that $f$ is close
to being a step function, $f(0;\phi)\approx\theta(\phi_{q}-\phi)$,
effectively cutting the integration at $\phi=\phi_{q}$. The relevant
value of $\phi_{q}$ will be determined from an explicit analytic
approximation for $f(0;\phi)$. 

If we use the number of $e$-foldings, $N\equiv\ln a$, as the time
variable $t$, the same method yields the average number of $e$-foldings,
$\left\langle \Delta N_{\text{NEC}}\right\rangle $, during the last
NEC-preserving portion of the trajectory before reheating. Below we
will compute $\phi_{q}$ and $\left\langle \Delta N_{\text{NEC}}\right\rangle $
explicitly for inflationary models with a power-law potential. In
a specific example, we will show that the average number of NEC-preserving
$e$-foldings in the model with $V(\phi)=\frac{1}{2}m^{2}\phi^{2}$
is of order $10^{5}$, if one assumes the model parameters that fit
the WMAP data. In this model, the value of $\phi_{q}$ turns out to
be such that $b^{2}(\phi_{q})\approx14$. Thus, $\phi_{q}$ is well
outside the fluctuation-dominated regime.

These results can be compared with the upper bound on the $e$-folds
of inflation obtained in Ref.~\cite{ArkaniHamed:2007ky}. Assuming
that the fluctuations are never dominant (equivalently, that the NEC
always holds), it was found that the number of observable $e$-folds
of inflation must be smaller than the entropy $S_{dS}$ of the final
de Sitter state \emph{after} inflation. The latter is an extremely
large number, of order $10^{120}$ (if we use the current value of
the dark energy density). The present calculation shows that the limit
on the number of observable $e$-folds is much more stringent. Therefore, 
the number of observable degrees of freedom, if it is expressed through
the entropy of the final de Sitter state, is in any case not directly
related to the total number of observable $e$-folds of inflation.

It must be noted that the value $\left\langle \Delta N_{\text{NEC}}\right\rangle $
is highly model-dependent. In the present paper, we perform computations
only for models with $V(\phi)\propto\phi^{2n}$ and derive a formula
for $\left\langle \Delta N_{\text{NEC}}\right\rangle $ {[}Eq.~(\ref{eq:DeltaN ans 01})]
that shows a sensitive dependence on $n$. However, the method of
calculation developed in this paper is sufficiently general so that
the number of total observable $e$-folds can be computed in any other
model of single-field slow-roll inflation.

\section{The time of the last NEC violation}

The evolution of the inflaton $\phi(t)$ for a model of type~(\ref{eq:inflaton action})
is a random process described by the Fokker-Planck (FP) equation~(see
e.g.~\cite{Linde:1993xx,Winitzki:2006rn})\begin{align}
\partial_{t}P(\phi,t) & =\hat{L}_{\phi}P(\phi,t),\label{eq:FP equ c1}\\
\hat{L}_{\phi}P & \equiv\partial_{\phi}\left(\partial_{\phi}(DP)-vP\right),\end{align}
where the coefficients $D(\phi)$ and $v(\phi)$ were defined above.
The FP equation is supplemented by appropriate initial and boundary
conditions. The initial condition \begin{equation}
P(\phi,t=0)=\delta(\phi-\phi_{\text{in}})\end{equation}
reflects the initial value of the inflaton field at $t=0$, while
the boundary conditions are imposed at the Planck boundary $\phi=\phi_{\text{Pl}}$
(we impose the reflecting boundary condition) and at the reheating
boundary $\phi=\phi_{*}$:\begin{equation}
\left[\partial_{\phi}(DP)-vP\right](\phi_{\text{Pl}},t)=0;\quad\partial_{\phi}(DP)(\phi_{*},t)=0.\label{eq:bc}\end{equation}
These equations describe the {}``comoving'' evolution, i.e.~$P(\phi,t)d\phi$
is the (infinitesimal) probability of having the value of the inflaton
within the interval $\left[\phi,\phi+d\phi\right]$ at a fixed point
in space and at time $t$. The {}``propagator'' (i.e.~Green's function)
$P(\phi_{1},\phi_{2},t)$ for the FP equation describes the probability
density of reaching the value $\phi=\phi_{2}$ at time $t$ starting
with $\phi=\phi_{1}$ at time $t=0$. The propagator is the solution
of Eq.~(\ref{eq:FP equ c1}) with respect to $\phi_{2}$ with the
initial condition \begin{equation}
P(\phi_{1},\phi_{2},0)=\delta(\phi_{1}-\phi_{2})\end{equation}
and the same boundary conditions as the FP equation, namely Eq.~(\ref{eq:bc}),
with respect to $\phi_{2}$. The propagator is a function only of
the time interval $t$ since the evolution is invariant under time
translations. The probabilistic interpretation of the propagator is
that $P(\phi_{1},\phi_{2},t)d\phi_{2}$ gives the probability of the
field value $\phi(t)$ being within the interval $[\phi_{2},\phi_{2}+d\phi_{2}]$
at time $t$, while $\phi_{1}\equiv\phi(0)$ and $t$ are sharply
fixed.

A technical complication for the present considerations is that an
equation describing the statistics of NEC-preserving evolution cannot
be formulated as another FP equation, e.g.~with modified coefficients.
This is so because the evolution strictly according to the FP equation
will violate the NEC at \emph{any} time. Mathematically, FP equations
describe a Brownian motion superimposed onto a deterministic motion,
while the Brownian motion admits large velocity fluctuations at short
time scales: A typical fluctuation $\delta\phi\propto\sqrt{\delta t}$
over a time $\delta t$ produces a velocity fluctuation $\propto{(\delta t)}^{-1/2}$,
which is unbounded as $\delta t\rightarrow0$. Thus, the description
of the inflaton through Brownian motion effectively excludes the possibility
that the trajectory $\phi(t)$ is strictly monotonic with $\dot{\phi}<0$
for \emph{any} finite duration of time. In reality, the mathematical
picture of Brownian motion does not hold for $\phi(t)$ at arbitrarily
small time scales. The FP equations may be used to describe the evolution
of $\phi$ only on time scales of order $\delta t\sim H^{-1}$ or
larger. A formulation of the stochastic evolution restricted to the
subset of NEC-preserving trajectories requires an averaging over such
time scales.

Therefore, one can formulate a statistical description of the subset
of NEC-preserving trajectories only by using an equation that is nonlocal
in time on time scales $\delta t$, or nonlocal in $\phi$ on some
relevant scale $\delta\phi$. Below we will derive one such equation
and obtain its approximate solution. For now, we focus on determining
the time $t_{\text{N}}$ of the last NEC violation, supposing that
the statistical distribution of NEC-preserving trajectories is known.

We consider the ensemble of comoving worldlines with inflaton trajectories
$\phi(t)$ starting at $t=0$ with a fixed value $\phi(0)\equiv\phi_{\text{in}}$.
The time $t_{\text{N}}$ is a random variable whose distribution can
be computed as follows. We ask for the probability $\text{Pr}\left(t_{\text{N}}<T\right)$
of the event $t_{\text{N}}<T$, where $T$ is a fixed parameter. The
event $t_{\text{N}}<T$ means that the NEC holds for $\phi(t)$ after
time $T$ and until reheating but may be violated at any earlier time
$t<T$. To compute $\text{Pr}\left(t_{\text{N}}<T\right)$, we split
the random trajectory $\phi(t)$ into two stages: The first stage
is the evolution from $\phi=\phi_{\text{in}}$ at $t=0$ to some intermediate
value $\phi_{T}$ at time $t=T$; during this first stage, the NEC
may be violated. The second stage is an NEC-preserving evolution from
$\phi=\phi_{T}$ at time $T$ until reheating at $\phi=\phi_{*}$
at some (random and not fixed) later time $t_{*}\geq T$. It is clear
that the last NEC violation happens before $t=T$ for any trajectory
consisting of these two stages, for any $\phi_{T}$ and $t_{*}$.
On the other hand, trajectories with different values of $\phi_{T}$
or $t_{*}$ are mutually exclusive random events. Therefore, we may
simply integrate over all allowed values of $\phi_{T}$ and $t_{*}$
in order to compute the probability $\text{Pr}\left(t_{\text{N}}<T\right)$.

We will now compute the probability of the event that the trajectory
$\phi(t)$ has the two stages as just described. Since the first stage
is not constrained with respect to the NEC, the evolution proceeds
according to the FP equation. The probability of reaching an intermediate
value $\phi_{T}$ at $t=T$ is thus given by the propagator $P(\phi_{\text{in}},\phi_{T},T)d\phi_{T}$. 

The evolution during the second stage needs to be NEC-preserving;
in the single-field model we are considering, this is synonymous with
the trajectory $\phi(t)$ being monotonic.

Since we are interested in the last NEC-preserving segment of the
trajectory $\phi(t)$ before reheating at $\phi=\phi_{*}$, we need
to compute the probability density of reaching a fixed value $\phi=\phi_{*}$
at an \emph{unknown} time $t_{*}$, rather than of reaching an unknown
value $\phi_{2}$ at a \emph{fixed} time $t$. Let us denote by $P_{+}(\phi_{0};t_{*})dt_{*}$
the probability of an NEC-preserving trajectory that starts at $\phi(t=0)=\phi_{0}$
and reaches $\phi=\phi_{*}$ within a time interval $[t_{*},t_{*}+dt_{*}]$.
(In this section, we will treat $P$ and $P_{+}$ as known; the necessary
computations are postponed to the next sections.)

Now we may express $\text{Pr}\left(t_{\text{N}}<T\right)$ as the
integral over $t_{*}$ and $\phi_{T}$ of the probability density
\begin{align}
 & \text{Pr}\,(t_{\text{N}}<T;t_{*},\phi_{T})dt_{*}d\phi_{T}\nonumber \\
 & \quad=P(\phi_{\text{in}},\phi_{T},T)d\phi_{T}P_{+}(\phi_{T},t_{*}-T)dt_{*},\label{eq:PRt0}\end{align}
namely,\begin{align}
 & \text{Pr}\left(t_{\text{N}}<T\right)=\negmedspace\int_{T}^{\infty}\negmedspace dt_{*}\negmedspace\int\negmedspace d\phi_{T}\,\text{Pr}\left(t_{\text{N}}<T;t_{*},\phi_{T}\right)\nonumber \\
 & \quad=\negmedspace\int_{T}^{\infty}\negmedspace dt_{*}\negmedspace\int\negmedspace d\phi_{T}\, P(\phi_{\text{in}},\phi_{T},T)P_{+}(\phi_{T};t_{*}-T).\label{eq:Prt1}\end{align}
 Here and below, the omitted range of integration over $\phi_{T}$
is from $\phi_{\text{Pl}}$ to $\phi_{*}$.

Once the probability $\text{Pr}\left(t_{\text{N}}<T\right)$ is known,
the probability density $p(t_{\text{N}})$ will be found from\begin{equation}
p(T)=\frac{\partial}{\partial T}\text{Pr}\left(t_{\text{N}}<T\right).\end{equation}
However, we will not proceed to compute $p(t_{\text{N}})$ since our
focus is on the duration $\Delta t\equiv t_{*}-t_{\text{N}}$ of the
last NEC-preserving segment of the trajectory $\phi(t)$. (As discussed
above, the quantity $\Delta t$ is observable in principle, while
$t_{\text{N}}$ is not observable.) 

Let us denote by $p(\Delta t)$ the probability density for $\Delta t$;
by definition $\Delta t\geq0$. It is more convenient to compute the
generating function\begin{equation}
g(\lambda)\equiv\negmedspace\int_{0}^{\infty}\negmedspace d\tau\, e^{\lambda\tau}p(\tau).\end{equation}
To assure the convergence of this integral, we will use $g(\lambda)$
only with $\lambda\leq0$. Once this function is known, we can compute
the moments of the distribution $p(\Delta t)$. For instance, the
mean value of $\Delta t$ and the dispersion $\sigma_{\Delta t}$
are given by\begin{align}
\left\langle \Delta t_{\text{NEC}}\right\rangle  & =g^{\prime}(0)\equiv\left.\frac{\partial g}{\partial\lambda}\right|_{\lambda=0};\\
\sigma_{\Delta t}^{2} & \equiv\left\langle \Delta t_{\text{NEC}}^{2}\right\rangle -\left\langle \Delta t_{\text{NEC}}\right\rangle ^{2}=g^{\prime\prime}(0)-g^{\prime2}(0).\label{eq:sigma def}\end{align}
The physical interpretation of $\left\langle \Delta t_{\text{NEC}}\right\rangle $
is the mean time spent in the last NEC-preserving segment of the trajectory
$\phi(t)$ before reheating, while $\sigma_{\Delta t}$ is the typical
deviation from the mean among all the trajectories $\phi(t)$.

In order to compute $g(\lambda)$, we consider the joint probability
density of the time of the last NEC violation with the parameters
$t_{*}$ and $\phi_{T}$; this probability density is found using
Eq.~(\ref{eq:PRt0}) as\begin{align}
p(T;t_{*},\phi_{T}) & \equiv\frac{\partial}{\partial T}\text{Pr}\,(t_{\text{N}}<T;t_{*},\phi_{T})\nonumber \\
 & =\frac{\partial}{\partial T}\left[P(\phi_{\text{in}},\phi_{T},T)P_{+}(\phi_{T},t_{*}-T)\right].\end{align}
The value of $g(\lambda)$ equals the average of $e^{\lambda\Delta t}=e^{\lambda(t_{*}-T)}$
among all trajectories that reheat at $t=t_{*}$ and contain the last
NEC violation at $t=T$. Hence \begin{align}
g(\lambda) & =\negmedspace\int d\phi_{T}\negmedspace\int_{0}^{\infty}\negmedspace dT\negmedspace\int_{T}^{\infty}\negmedspace dt_{*}e^{\lambda(t_{*}-T)}p(T;t_{*},\phi_{T})\nonumber \\
 & =\negmedspace\int d\phi_{T}\negmedspace\int_{0}^{\infty}\negmedspace dt_{*}\negmedspace\int_{0}^{t_{*}}\negmedspace dT\times\nonumber \\
 & \qquad\times e^{\lambda(t_{*}-T)}\frac{\partial}{\partial T}\left[P(\phi_{\text{in}},\phi_{T},T)P_{+}(\phi_{T},t_{*}-T)\right].\label{eq:glambda int}\end{align}
Integrating by parts, we obtain\begin{align*}
 & \;\int_{0}^{t_{*}}\negmedspace dTe^{\lambda(t_{*}-T)}\frac{\partial}{\partial T}\left[P(\phi_{\text{in}},\phi_{T},T)P_{+}(\phi_{T},t_{*}-T)\right]\\
 & =P(\phi_{\text{in}},\phi_{T},t_{*})P_{+}(\phi_{T},0)-e^{\lambda t_{*}}P(\phi_{\text{in}},\phi_{T},0)P_{+}(\phi_{T},t_{*})\\
 & \quad+\lambda\int_{0}^{t_{*}}\negmedspace dTe^{\lambda(t_{*}-T)}P(\phi_{\text{in}},\phi_{T},T)P_{+}(\phi_{T},t_{*}-T)\\
 & =P(\phi_{\text{in}},\phi_{T},t_{*})\delta(\phi_{T}-\phi_{*})-e^{\lambda t_{*}}\delta(\phi_{T}-\phi_{\text{in}})P_{+}(\phi_{T},t_{*})\\
 & \quad+\lambda\int_{0}^{t_{*}}\negmedspace dTe^{\lambda(t_{*}-T)}P(\phi_{\text{in}},\phi_{T},T)P_{+}(\phi_{T},t_{*}-T).\end{align*}
Substituting this into Eq.~(\ref{eq:glambda int}) and simplifying,
we find\begin{align}
g(\lambda) & =\negmedspace\int_{0}^{\infty}\negmedspace dt_{*}\negmedspace\left[P(\phi_{\text{in}},\phi_{*},t_{*})-e^{\lambda t_{*}}P_{+}(\phi_{\text{in}},t_{*})\right]\nonumber \\
 & \;+\lambda\int d\phi_{T}\negmedspace\int_{0}^{\infty}\negmedspace dT\negmedspace\int_{0}^{\infty}\negmedspace d(t_{*}-T)\times\nonumber \\
 & \qquad\times e^{\lambda(t_{*}-T)}P(\phi_{\text{in}},\phi_{T},T)P_{+}(\phi_{T},t_{*}-T)\nonumber \\
 & =1-f(\lambda;\phi_{\text{in}})+\lambda\int d\phi_{T}\Psi(\phi_{\text{in}},\phi_{T})f(\lambda;\phi_{T}),\label{eq:g lambda}\end{align}
where we defined the auxiliary functions\begin{align}
f(\lambda;\phi) & \equiv\int_{0}^{\infty}d\tau\, e^{\lambda\tau}P_{+}(\phi,\tau);\\
\Psi(\phi_{\text{in}},\phi) & \equiv\int_{0}^{\infty}dT\, P(\phi_{\text{in}},\phi,T).\end{align}
So we will not actually need explicit expressions for the full distributions
$P(\phi_{\text{in}},\phi,T)$ and $P_{+}(\phi,\tau)$; it suffices
to compute the functions $f$ and $\Psi$. 

As shown in Eq.~(\ref{eq:psi approx ans}) in Appendix~\ref{sub:Solving-stationary-FP}
below, the function $\Psi(\phi_{\text{in}},\phi)$ can be approximated
(up to slow-roll corrections) for $\phi<\phi_{\text{in}}$ by\begin{equation}
\Psi(\phi_{\text{in}},\phi)\approx\frac{1}{-v(\phi)}.\end{equation}
The dependence on the value of $\phi_{\text{in}}$ was omitted here
because it is exponentially small as long as $\phi_{\text{in}}$ is
within the diffusion-dominated regime. Thus we will omit the dependence
on $\phi_{\text{in}}$ where appropriate. 

The function $f(\lambda;\phi)$ will be computed in Sec.~\ref{sub:Propagator-for-NEC-preserving}
as\begin{equation}
f(\lambda;\phi_{T})=\exp\left[-\int_{\phi_{*}}^{\phi_{T}}W(\lambda;\phi)d\phi\right],\label{eq:W def}\end{equation}
where the auxiliary function $W(\lambda;\phi)$ is approximately found
as the solution of Eq.~(\ref{eq:W equ 2}) below. We can then rewrite
Eq.~(\ref{eq:g lambda}) as\begin{equation}
g(\lambda)=1-f(\lambda;\phi_{\text{in}})+\lambda\negmedspace\int_{\phi_{*}}^{\phi_{\text{Pl}}}\negmedspace\frac{d\phi}{-v(\phi)}f(\lambda;\phi).\end{equation}
We note that $f(0;\phi_{T})$ is interpreted physically as the total
probability of never violating the NEC for a trajectory starting at
$\phi=\phi_{T}$. If $\phi_{\text{in}}$ is in the diffusion-dominated
regime, the probability $f(0;\phi_{\text{in}})$ is exponentially
small and can be neglected in Eq.~(\ref{eq:g lambda}). It follows
that\begin{equation}
\left\langle \Delta t_{\text{NEC}}\right\rangle =g^{\prime}(0)\approx\negmedspace\int_{\phi_{*}}^{\phi_{\text{Pl}}}\negmedspace\frac{d\phi}{-v(\phi)}f(0;\phi).\label{eq:tau NEC}\end{equation}
In the rest of the paper we will perform the calculations explicitly
and show that the factor $f(0;\phi)$ effectively cuts off the integration
at a model-dependent value $\phi=\phi_{q}$, which is in the regime
where the diffusion is already small. A numerical calculation in a
specific model of inflation is then given in Sec.~\ref{sub:Example:-Inflation-with}.

\section{Duration of NEC-preserving trajectories\label{sub:Propagator-for-NEC-preserving}}

It is necessary for our purposes to compute the function in Eq.~(\ref{eq:W def}),
which we denoted by $f$: \begin{equation}
f(\lambda;\phi_{0})\equiv\int_{0}^{\infty}\negmedspace d\tau\, e^{\lambda\tau}P_{+}(\phi_{0},\tau).\label{eq:def f}\end{equation}
This is the generating function of the duration $\tau$ of NEC-preserving
trajectories starting at a given value $\phi=\phi_{0}$ at time $t=0$
and finishing at $\phi=\phi_{*}$ at an unknown time $t=\tau$. For
instance, the mean duration of time until reheating among all the
NEC-\emph{preserving} trajectories starting at $\phi=\phi_{0}$ is
given by\begin{equation}
\left\langle \tau\right\rangle =\left.\frac{\partial}{\partial\lambda}\ln f(\lambda;\phi_{0})\right|_{\lambda=0}.\end{equation}

As discussed above, we expect that the function $f(\lambda;\phi_{0})$
satisfies an equation nonlocal in $\phi_{0}$. To derive this equation,
we consider the change $\delta\phi=\phi(\delta t)-\phi(0)$ of the
(spatially coarse-grained) value of $\phi$ after a single Hubble
time step $\delta t\equiv H^{-1}$ at a given comoving point in space,\begin{equation}
\delta\phi(\phi,\xi)\equiv v(\phi)\delta t+\xi\sqrt{2D(\phi)\delta t},\end{equation}
where $\xi$ is a normally distributed random variable. We denote
for convenience by $p(\xi)$ the probability density of $\xi$,\begin{equation}
p(\xi)\equiv\frac{1}{\sqrt{2\pi}}\exp\left(-{\textstyle \frac{1}{2}}\xi^{2}\right).\end{equation}
The NEC-preserving property at the presently considered Hubble time
step is equivalent to the condition $\delta\phi<0$ or\begin{equation}
\xi<b(\phi)\equiv\frac{-v(\phi)\delta t}{\sqrt{2D(\phi)\delta t}}=\frac{H^{\prime}M_{\text{Pl}}^{2}}{2H^{2}}.\label{eq:b def}\end{equation}
(Note that the quantity $b$ is always positive since $H^{\prime}=dH/d\phi>0$
due to the assumption $dV/d\phi>0$.) Since the probability $P_{+}(\phi_{0},\tau)$
includes only trajectories that preserve NEC throughout their evolution,
we must include only values of $\xi$ such that $\xi<b(\phi_{0})$
when we describe the Hubble time step leading from $\phi_{0}$ to
$\phi_{0}+\delta\phi$. So we may express $P_{+}(\phi_{0},\tau)$
through $P_{+}(\phi_{0}+\delta\phi,\tau-\delta t)$ as\begin{equation}
P_{+}(\phi_{0};\tau)=\negmedspace\int_{-\infty}^{b(\phi_{0})}\negmedspace d\xi\, p(\xi)\, P_{+}(\phi_{0}+\delta\phi,\tau-\delta t).\label{eq:P plus 2}\end{equation}
Here $\delta\phi\equiv\delta\phi(\phi_{0},\xi)$ under the integral
is understood as a function of $\xi$. Using Eq.~(\ref{eq:def f}),
we may now express the value $f(\lambda;\phi_{0})$ through the values
of $f$ at the next Hubble step as follows. We first integrate Eq.~(\ref{eq:P plus 2})
with $e^{\lambda\tau}d\tau$ from $\tau=\delta t$ to infinity and
then exchange the order of integrals and shift the integration variable
$\tau$ by $\delta t$: \begin{align}
 & \negmedspace\int_{\delta t}^{\infty}\negmedspace d\tau\, e^{\lambda\tau}P_{+}(\phi_{0},\tau)\nonumber \\
 & \;=\negmedspace\int_{\delta t}^{\infty}\negmedspace d\tau\, e^{\lambda\tau}\negmedspace\int_{-\infty}^{b(\phi_{0})}\negmedspace d\xi\, p(\xi)\, P_{+}(\phi_{0}+\delta\phi,\tau-\delta t)\nonumber \\
 & \;=\negmedspace\int_{-\infty}^{b(\phi_{0})}\negmedspace d\xi\, p(\xi)\negmedspace\int_{0}^{\infty}\negmedspace d\tau\, e^{\lambda(\tau+\delta\tau)}\, P_{+}(\phi_{0}+\delta\phi,\tau)\nonumber \\
 & \;=\negmedspace\int_{-\infty}^{b(\phi_{0})}\negmedspace d\xi\, p(\xi)\, e^{\lambda\delta t}f(\lambda;\phi_{0}+\delta\phi).\label{eq:P plus 3}\end{align}
Note that the top line in Eq.~(\ref{eq:P plus 3}) is slightly different
from the definition of $f(\lambda;\phi_{0})$: The integration proceeds
from $\tau=\delta t$ rather than from $\tau=0$ in order to allow
the subtraction $\tau-\delta t$ in the argument of $P_{+}$. The
difference, \begin{equation}
\int_{0}^{\delta t}\negmedspace d\tau\, P_{+}(\phi_{0},\tau)\approx\delta t\, P_{+}(\phi_{0},\delta t),\end{equation}
is negligible as long as $\phi_{0}$ is at least a few $e$-foldings
away from reheating. This is so because $P_{+}(\phi_{0},\delta t)$
is equal to the (exponentially small) probability of jumping from
$\phi=\phi_{0}$ directly to $\phi=\phi_{*}$ in one Hubble time $\delta t$.
Therefore, we may replace the top line in Eq.~(\ref{eq:P plus 3})
by $f(\lambda;\phi_{0})$ and finally obtain the equation\begin{equation}
f(\lambda;\phi_{0})=\negmedspace\int_{-\infty}^{b(\phi_{0})}\negmedspace d\xi\, p(\xi)\, e^{\lambda\delta t}f(\lambda;\phi_{0}+\delta\phi(\phi_{0},\xi)).\label{eq:f equ 1}\end{equation}
This is the basic equation describing the function $f(\lambda;\phi_{0})$;
as expected, it is nonlocal in $\phi$.

It is not possible to approximate Eq.~(\ref{eq:f equ 1}) by a diffusion
equation (as is the normal procedure while deriving FP equations)
because the integration in Eq.~(\ref{eq:f equ 1}) proceeds over
a $\phi$-dependent range. Rather than trying to solve Eq.~(\ref{eq:f equ 1})
directly, we will approximate the solution of Eq.~(\ref{eq:f equ 1})
by an adiabatic ansatz {[}Eq.~(\ref{eq:f ansatz}) below]. 

Up to now we have been using the proper time as the variable $t$.
If a different time parameterization is desired, such as \begin{equation}
\tilde{t}=\int^{t}A(\phi(t))dt,\end{equation}
where $A(\phi)$ is a known function, then the coefficients $D$,
$v$, and $\delta t$ must be modified as follows,\begin{equation}
\tilde{v}=\frac{v}{A},\quad\tilde{D}=\frac{D}{A},\quad\delta\tilde{t}=A\delta t,\end{equation}
while the dimensionless coefficient $b(\phi)$ is unchanged. For instance,
passing to the $e$-folding time \begin{equation}
N=\int^{t}Hdt\end{equation}
is implemented by choosing $A(\phi)=H(\phi)$. Below we will compute
$\left\langle \Delta N_{\text{NEC}}\right\rangle $ in a specific
model of inflation by using this method.

At the end of the calculation, we will only need to evaluate $f(\lambda=0;\phi_{0})$.
As already mentioned above, $f(0;\phi_{0})$ is the fraction of trajectories
that never violate the NEC among all trajectories $\phi(t)$ starting
at $\phi=\phi_{0}$. We note that for $\phi_{0}$ in the fluctuation-dominated
regime, the probability $f(0;\phi_{0})$ rapidly decreases with growing
$\phi_{0}$ because there is a significant probability of violating
the NEC at every Hubble time step at those $\phi_{0}$. On the other
hand, the probability of violating the NEC in the no-diffusion regime
is exponentially small, and hence $f(0;\phi_{0})$ is nearly constant
and almost equal to 1 for $\phi_{0}$ in that regime. Therefore, we
expect that $f(\lambda;\phi_{0})$ has exponentially strong dependence
on $\phi_{0}$. Moreover, $P_{+}(\phi_{*},\tau)=\delta(\tau)$; this
can be shown by considering\begin{align}
\int_{0}^{\infty}P_{+}(\phi_{*},\tau)d\tau & =1,\\
P_{+}(\phi_{*},\tau) & =0\:\text{for}\:\tau>0,\end{align}
which holds because trajectories starting with $\phi=\phi_{*}$ immediately
reheat and have zero duration. Therefore \begin{equation}
f(\lambda;\phi_{*})=\int_{0}^{\infty}e^{\lambda\tau}P_{+}(\phi_{*},\tau)d\tau=1.\label{eq:f lambda phi bc}\end{equation}
Motivated by these considerations, we represent the exponential behavior
of $f(\lambda;\phi_{0})$ and the boundary condition~(\ref{eq:f lambda phi bc})
by the ansatz \begin{equation}
f(\lambda;\phi_{0})=\exp\left[-\int_{\phi_{*}}^{\phi_{0}}\negmedspace W(\lambda;\phi)d\phi\right],\label{eq:f ansatz}\end{equation}
where $W(\lambda;\phi)$ is a new unknown function such that $W(0;\phi)>0$.
We then divide Eq.~(\ref{eq:f equ 1}) through by $f(\lambda;\phi_{0})$
and expand $f(\lambda;\phi+\delta\phi)$ to first order in $\delta\phi$:\begin{align}
1 & =\negmedspace\int_{-\infty}^{b(\phi_{0})}\negmedspace d\xi\, p(\xi)\, e^{\lambda\delta t}\exp\left[-\int_{\phi_{0}}^{\phi_{0}+\delta\phi}\negmedspace W(\lambda;\phi)d\phi\right]\nonumber \\
 & \approx\negmedspace\int_{-\infty}^{b(\phi_{0})}\negmedspace d\xi\, p(\xi)\, e^{\lambda\delta t}\exp\left[-W(\lambda;\phi_{0})\delta\phi(\phi_{0},\xi)\right]\nonumber \\
 & =e^{\left(\lambda-vW\right)\delta t}\negmedspace\int_{-\infty}^{b}\negthickspace\frac{d\xi}{\sqrt{2\pi}}\exp\left[-\frac{\xi^{2}}{2}-\sqrt{2D\delta t}W\xi\right]\nonumber \\
 & =e^{\left(\lambda-vW+DW^{2}\right)\delta t}\left[\frac{1}{2}+\frac{1}{2}\text{erf}\left(\frac{b+W\sqrt{2D\delta t}}{\sqrt{2}}\right)\right],\label{eq:W equ 1}\end{align}
where we suppressed the argument $\phi_{0}$ in the last line, $b\equiv b(\phi)$
was defined in Eq.~(\ref{eq:b def}), while $\text{erf}\, x$ is
the standard error function\begin{equation}
\text{erf }x\equiv\frac{2}{\sqrt{\pi}}\int_{0}^{x}e^{-t^{2}}dt.\end{equation}

We now note that Eq.~(\ref{eq:W equ 1}) does not contain derivatives
of $W$; this means that we are using an adiabatic approximation where
$W(\lambda;\phi)$ is assumed to vary slowly with $\phi$. Thus $W(\lambda;\phi)$
is the unique real root of the transcendental equation\begin{equation}
\exp\left[\left(vW-DW^{2}-\lambda\right)\delta t\right]=\frac{1}{2}+\frac{1}{2}\text{erf}\left[\frac{b+W\sqrt{2D\delta t}}{\sqrt{2}}\right]\label{eq:W equ 2}\end{equation}
such that $W(\lambda;\phi)>0$ for $\lambda=0$. 

The coefficient $b(\phi)$ measures the influence of quantum fluctuations
on the evolution $\phi(t)$. It is possible to obtain approximate
solutions of Eq.~(\ref{eq:W equ 2}) in the cases $b\gg1$ (a nearly
fluctuation-free regime) and $b\ll1$ (a fluctuation-dominated regime).
To simplify calculations, we pass to a new dimensionless variable
$r$ by rewriting Eq.~(\ref{eq:W equ 2}) as \begin{equation}
\frac{r^{2}}{2}+\ln\left[\frac{1}{2}+\frac{1}{2}\text{erf}\frac{r}{\sqrt{2}}\right]\equiv L(r)=\frac{b^{2}}{2}-\lambda\delta t,\label{eq:b equ 1}\end{equation}
where $r(\lambda;\phi)$ is related to $W(\lambda;\phi)$ by \begin{equation}
W(\lambda;\phi)\equiv\frac{r(\lambda;\phi)-b}{\sqrt{2D\delta t}}.\label{eq:W thru r}\end{equation}
We solve Eq.~(\ref{eq:b equ 1}) by using the inverse function $L^{-1}(x)$,\begin{equation}
r=L^{-1}({\textstyle \frac{1}{2}b^{2}-\lambda\delta t)},\end{equation}
which then gives the solution $W(\lambda;\phi)$ through Eq.~(\ref{eq:W thru r})
as\begin{equation}
W(\lambda;\phi)=\frac{L^{-1}(\frac{1}{2}b^{2}-\lambda\delta t)-b}{\sqrt{2D\delta t}}.\label{eq:W thru Linv}\end{equation}

Derivatives of $W(\lambda;\phi)$ with respect to $\lambda$ can be
expressed through $W(\lambda;\phi)$ by computing the derivative of
$L^{-1}$,\begin{align}
L^{\prime}(r) & =r+\sqrt{\frac{2}{\pi}}\frac{e^{-\frac{1}{2}r^{2}}}{1+\text{erf}\frac{r}{\sqrt{2}}}=r+\frac{e^{-L(r)}}{\sqrt{2\pi}},\\
\frac{\partial}{\partial x}\left[L^{-1}(x)\right] & =\frac{1}{L^{\prime}[L^{-1}(x)]}=\left[L^{-1}(x)+\frac{e^{-x}}{\sqrt{2\pi}}\right]^{-1},\end{align}
 and using Eq.~(\ref{eq:W thru Linv}). For instance, we find\begin{align}
\frac{\partial}{\partial\lambda}W & =-\frac{\delta t}{\sqrt{2D\delta t}}\frac{1}{L^{\prime}[L^{-1}(\frac{1}{2}b^{2}-\lambda\delta t)]}\nonumber \\
 & ={\frac{1}{v}\left[1-\frac{2D}{v}W+\frac{e^{-\frac{1}{2}b^{2}+\lambda\delta t}}{b\sqrt{2\pi}}\right]}^{-1}.\label{eq:dW def}\end{align}
Further derivatives with respect to $\lambda$ can be obtained similarly.
Since we will ultimately compute the generating function $g(\lambda)$
and its derivatives only at $\lambda=0$, it is sufficient to set
$\lambda=0$ in what follows.

We also note that the function $L(r)$ is {}``universal'' in the
sense that its definition does not depend on the inflaton potential
$V(\phi)$. It is therefore useful to approximate the inverse function
$L^{-1}(x)$ semi-numerically. We first consider the asymptotic behavior
of $L^{-1}(x)$ for large $x$. Using the well-known asymptotic representation
of the error function,\begin{equation}
\text{erf }x=1-\frac{1}{x\sqrt{\pi}}e^{-x^{2}}\left(1+O(x^{-2})\right),\quad x\rightarrow+\infty,\end{equation}
we obtain\begin{equation}
L(r)=\frac{r^{2}}{2}-\frac{\exp\left(-\frac{1}{2}r^{2}\right)}{r\sqrt{2\pi}}\left[1+O(r^{-2})\right],\; r\rightarrow+\infty,\end{equation}
and hence\begin{equation}
L^{-1}(x)=\sqrt{2x}+\frac{e^{-x}}{2x\sqrt{2\pi}}\left[1+O(x^{-1})\right],\; x\rightarrow+\infty.\end{equation}
This asymptotic formula allows us to obtain the approximate solution
$W(0;\phi)$ for the case $b\gg1$ as\begin{equation}
W(0;\phi)=\frac{L^{-1}(\frac{1}{2}b^{2})-b}{\sqrt{2D\delta t}}\approx\frac{1+O(b^{-2})}{\sqrt{2\pi}}\frac{e^{-\frac{1}{2}b^{2}}}{b^{2}\sqrt{2D\delta t}}.\label{eq:W approx large}\end{equation}

The solution in the opposite regime $b\ll1$ can be found by starting
with the numerically obtained value \begin{equation}
L^{-1}(0)\approx0.7286\equiv r_{0}\end{equation}
and by expanding $L^{-1}(x)$ near $x=0$,\begin{align}
L^{-1}(x) & =r_{0}+x\left.\frac{\partial}{\partial x}\right|_{x=0}\left[L^{-1}\right]+O(x^{2})\nonumber \\
 & =r_{0}+x\left[r_{0}+\frac{1}{\sqrt{2\pi}}\right]^{-1}+O(x^{2}),\; x\rightarrow0.\end{align}
Hence for $b\ll1$ we have\begin{equation}
W(0;\phi)=\frac{r_{0}-b+\frac{1}{2}b^{2}\left[r_{0}+\frac{1}{\sqrt{2\pi}}\right]^{-1}+O(b^{4})}{\sqrt{2D\delta t}},\; b\rightarrow0.\end{equation}
We have thus obtained the solution  $W(0;\phi)$ in the two opposite
regimes. An approximation that holds uniformly for all positive $b$
can be obtained, if desired, by matching the asymptotic expressions
near $b=0$ and $b=\infty$, for instance, using the following interpolating
function,\begin{equation}
W(0;\phi)\approx\frac{e^{-\frac{1}{2}b^{2}}}{\sqrt{2D\delta t}}\frac{b+0.7194}{b^{3}\sqrt{2\pi}+1.803b^{2}+2.728b+0.9874}.\end{equation}
Numerical verification shows that this function approximates $W(0,\phi)$
to within about 2.5\% relative precision for all $b>0$. (We note
that $W$ is model-independent only as a function of $b$ and $\sqrt{2D\delta t}$,
while $b(\phi)$, $D(\phi)$, and $\delta t\equiv H^{-1}(\phi)$ of
course depend on the chosen model of inflation.)

However, it turns out that the approximation in Eq.~(\ref{eq:W approx large}),
which holds in the fluctuation-free regime, is sufficient for our
present purposes. Let us derive the corresponding approximation for
the function $f(0;\phi_{0})$,\begin{equation}
f(0;\phi_{0})=\exp\left[-\int_{\phi_{*}}^{\phi_{0}}W(0;\phi)d\phi\right],\end{equation}
assuming that $\phi_{0}$ is such that $b^{2}(\phi_{0})\gg1$. Since
$W(0;\phi)$ is quickly growing with $\phi$, the integral under the
exponential above is dominated by the upper limit, so we can use the
asymptotic estimate\begin{align}
\negmedspace\int_{\phi_{*}}^{\phi_{0}}\negmedspace W(0;\phi)d\phi & \approx\negmedspace\int_{\phi_{*}}^{\phi_{0}}\negmedspace\frac{2\pi}{H}d\phi\frac{e^{-\frac{1}{2}b^{2}}}{b^{2}\sqrt{2\pi}}\left[1+O(b^{-2})\right]\nonumber \\
 & \approx\left.\frac{\sqrt{2\pi}}{H}\frac{e^{-\frac{1}{2}b^{2}}}{b^{3}}\left[-\frac{\partial b}{\partial\phi}\right]^{-1}\right|_{\phi=\phi_{0}}.\end{align}
In deriving this estimate, we neglected terms of order $b^{-2}$ as
well as derivatives of $H$ and $b$, since these are merely slow-roll
corrections. 

For $\phi_{0}$ near reheating, we have $f(0;\phi_{0})\approx1$ with
exponential precision. The value of $\phi_{q}$ at which $f(0;\phi_{q})$
first drops to $\exp(-1)$ can then be found as the solution of the
equation\begin{equation}
\frac{\sqrt{2\pi}}{H(\phi_{q})}\frac{e^{-\frac{1}{2}b_{q}^{2}}}{b_{q}^{3}}\left[-\frac{\partial b}{\partial\phi}\right]_{\phi=\phi_{q}}^{-1}=1,\label{eq:b q equ}\end{equation}
where we need to substitute $b_{q}\equiv b(\phi_{q})$. This can be
interpreted as a closed-form equation for $b_{q}$ if we express $\partial b/\partial\phi$
and $H(\phi_{q})$ as functions of $b_{q}$. A numerical calculation
needs to be performed to solve this equation for $b_{q}$ in a particular
inflationary model and to check that the resulting value of $b_{q}$
satisfies $b_{q}^{2}\gg1$, which is required for the validity of
the approximation used to derive Eq.~(\ref{eq:b q equ}). (For instance,
the calculations in the next section show that $b_{q}^{2}\approx14\gg1$
for the inflationary model with the potential $V(\phi)\propto\phi^{2}$.)

We have thus determined $\phi_{q}$ such that $f(0;\phi_{q})=e^{-1}$.
The function $f(0;\phi)$ has a sharp dependence on $\phi$ and interpolates
from 1 to 0 within a narrow interval around $\phi=\phi_{q}$. To estimate
the width of this interval, let us find the value $\phi_{q}^{(2)}$
such that $f(0;\phi_{q}^{(2)})=e^{-2}$. This value can be determined
as a solution of\begin{equation}
\frac{\sqrt{2\pi}}{H(\phi_{q}^{(2)})}\frac{e^{-\frac{1}{2}b^{2}}}{b^{3}}\left[-\frac{\partial b}{\partial\phi}\right]_{\phi=\phi_{q}^{(2)}}^{-1}=2.\end{equation}
Then the width of the interval in $b_{q}$ can be estimated as $\delta b_{q}=b_{q}^{(2)}-b_{q}$.
Since the exponential above is the fastest-varying function of $b$,
to first approximation we have\begin{equation}
b^{2}(\phi_{q}^{(2)})\approx b^{2}(\phi_{q})-2\ln2.\end{equation}
(Numerical calculations show that this is an overestimate of $\delta b$
by about 20\%.) This leads to a change in the value of $\phi_{q}$
that can be computed through\begin{equation}
\frac{\delta\phi_{q}}{\phi_{q}}\approx\frac{\delta b}{b}\left[\frac{\partial\ln b}{\partial\ln\phi}\right]^{-1},\quad\frac{\delta b}{b}\approx-b^{-2}\ln2.\end{equation}
Since $b^{-2}\ll1$ while the logarithmic derivative $\partial\ln b/\partial\ln\phi$
is not large, we find that $\delta\phi_{q}/\phi_{q}\ll1$. Hence,
the function $f(0;\phi_{0})$ has the effect of a cutoff near $\phi=\phi_{q}$
when integrated with a slowly-varying function of $\phi$ such as
$1/v(\phi)$, as required for Eq.~(\ref{eq:tau NEC}). 

Below we will replace integrations with the factor $f(0;\phi)$ by
integrations with the upper limit $\phi=\phi_{q}$. This approximation
introduces a certain error; to estimate the effect of this error on
the calculation of $\left\langle \Delta N_{\text{NEC}}\right\rangle $,
let us find the number of $e$-folds in the slow-roll trajectory between
$\phi=\phi_{q}^{(2)}$ and $\phi=\phi_{q}$:\begin{align}
\delta N_{q} & =\int_{\phi_{q}}^{\phi_{q}^{(2)}}\frac{H(\phi)d\phi}{-v(\phi)}\approx\frac{H(\phi_{q})\delta\phi_{q}}{-v(\phi_{q})}\nonumber \\
 & \approx\left.\frac{H\delta b}{-bv}\left[\frac{d\ln b}{d\phi}\right]^{-1}\right|_{\phi=\phi_{q}}=\left.\frac{H\ln2}{b^{2}v}\left[\frac{d\ln b}{d\phi}\right]^{-1}\right|_{\phi=\phi_{q}}.\end{align}
To estimate $d\ln b/d\phi$, we use Eqs.~(\ref{eq:b def 0}) and
find \begin{equation}
\frac{d\ln b}{d\phi}=\frac{v^{\prime}}{v}-\frac{1}{2}\frac{D^{\prime}}{D}-\frac{1}{2}\frac{H^{\prime}}{H}=-\frac{\varepsilon_{2}\sqrt{\pi}}{M_{\text{Pl}}\sqrt{\varepsilon_{1}}},\end{equation}
where $\varepsilon_{2}$ is the second slow-roll parameter,\begin{equation}
\varepsilon_{2}\equiv\frac{M_{\text{Pl}}^{2}}{4\pi}\left(\frac{V^{\prime2}}{V^{2}}-\frac{V^{\prime\prime}}{V}\right).\end{equation}
We also have\begin{equation}
\frac{-v}{H}=\frac{\sqrt{\pi}}{2}M_{\text{Pl}}\sqrt{\varepsilon_{1}}.\end{equation}
Therefore we obtain the estimate\begin{equation}
\delta N_{q}=\left.\frac{\ln4}{\pi b^{2}\varepsilon_{2}}\right|_{\phi=\phi_{q}}.\label{eq:delta Nq}\end{equation}
This estimate is important because it displays the error inherent
in the definition of the boundary of the fluctuation-dominated regime.
Below we will check that this error is acceptable when determining
the average number of observable $e$-folds. 

Let us summarize the calculations presented so far. We have derived
an estimate of the mean time $\left\langle \Delta t_{\text{NEC}}\right\rangle $
of the last NEC-preserving portion of the inflationary trajectory:\begin{equation}
\left\langle \Delta t_{\text{NEC}}\right\rangle \approx\int_{\phi_{*}}^{\phi_{\text{Pl}}}\frac{d\phi}{-v(\phi)}f(0;\phi)\approx\int_{\phi_{*}}^{\phi_{q}}\frac{d\phi}{-v(\phi)},\end{equation}
where the value of $\phi_{q}$ is determined from Eq.~(\ref{eq:b q equ}).
We have introduced a simple approximation where the function $f(0;\phi)$
is replaced by a step cut-off at $\phi=\phi_{q}$; the error of this
approximation is expected to be small. It is possible, in principle,
to determine the function $f(0;\phi)$ numerically and thus to obtain
a sharper estimate, as well as to compute the standard deviation $\sigma_{\Delta N}$
of the number of observable $e$-folds using Eq.~(\ref{eq:sigma def}).
However, we expect that the standard deviation $\sigma_{\Delta N}$
will not be larger than the width of the function $f(0;\phi)$ around
the point $\phi=\phi_{q}$, which is of order $\delta\phi_{q}$ as
estimated above. Therefore, it will be sufficient for the present
purposes to use the estimated width of the function $f(0;\phi)$ as
the statistical uncertainty in $\left\langle \Delta t_{\text{NEC}}\right\rangle $.

\section{Example: Inflation with a power-law potential\label{sub:Example:-Inflation-with}}

We now perform specific calculations of $\phi_{q}$ and the average
number of NEC-preserving $e$-folds, $\left\langle \Delta N_{\text{NEC}}\right\rangle $,
for a model of single-field inflation of type~(\ref{eq:inflaton action})
with the potential \begin{equation}
V(\phi)=\lambda M_{\text{Pl}}^{4}\left(\frac{\phi}{M_{\text{Pl}}}\right)^{2n}.\end{equation}
This model can fit the current observations when $n=1$ or $n=2$
(see, e.g., \cite{Leach:2003us}). We use the formalism developed
in the previous sections for computing $\left\langle \Delta t_{\text{NEC}}\right\rangle $,
except that we divide $\delta t$, $v(\phi)$, and $D(\phi)$ in every
formula by the factor $H(\phi)$ in order to pass from the proper
time $t$ to the $e$-folding time $N$.

For this model, we find in the slow-roll approximation\begin{align}
H(\phi) & =\sqrt{\frac{8\pi}{3}}M_{\text{Pl}}\sqrt{\lambda}\left(\frac{\phi}{M_{\text{Pl}}}\right)^{n},\label{eq:H phi 2 d}\\
v(\phi) & =-\frac{M_{\text{Pl}}^{2}}{4\pi}\sqrt{\frac{8\pi}{3}}n\sqrt{\lambda}\left(\frac{\phi}{M_{\text{Pl}}}\right)^{n-1},\\
b(\phi) & =-\frac{2\pi v}{H^{2}}=\frac{n}{\sqrt{\lambda}}\sqrt{\frac{3}{32\pi}}\left(\frac{M_{\text{Pl}}}{\phi}\right)^{n+1}.\label{eq:b phi 2 d}\end{align}
The slow-roll parameters (computed as functions of $\phi$ through
the potential $V$) are \begin{align}
\varepsilon_{1} & \equiv\frac{M_{\text{Pl}}^{2}}{16\pi}\frac{V^{\prime2}}{V^{2}}=\frac{n^{2}}{4\pi}\frac{M_{\text{Pl}}^{2}}{\phi^{2}};\\
\varepsilon_{2} & \equiv\frac{M_{\text{Pl}}^{2}}{4\pi}\left(\frac{V^{\prime2}}{V^{2}}-\frac{V^{\prime\prime}}{V}\right)=\frac{n}{2\pi}\frac{M_{\text{Pl}}^{2}}{\phi^{2}}.\end{align}
Reheating is assumed to happen at $\phi=\phi_{*}$ with $\varepsilon_{1}(\phi_{*})=1$,
which gives%
\footnote{The value of $\phi_{*}$ is only an estimate because it is computed
in the slow-roll approximation, which does not hold near reheating.
However, our results are not sensitive to the precise value of $\phi_{*}$.%
} \begin{equation}
\phi_{*}=M_{\text{Pl}}\frac{n}{\sqrt{4\pi}}.\end{equation}
The number $N_{e}$ of inflationary $e$-foldings accumulated between
some value $\phi=\phi_{1}$ until reheating is estimated (assuming
$\phi_{1}\gg\phi_{*}$) as\begin{equation}
N_{e}(\phi_{1})=\int_{\phi_{*}}^{\phi_{1}}\frac{H(\phi)d\phi}{-v(\phi)}\approx\frac{2\pi\phi_{1}^{2}}{nM_{\text{Pl}}^{2}}.\end{equation}
The squared amplitude of scalar primordial perturbations generated
at $\phi=\phi_{1}$ is given by WMAP observations as $P_{S}\approx2.3\cdot10^{-9}$
(we use the data from Ref.~\cite{Leach:2003us}). Assuming that
this amplitude is generated at $N_{e}$ $e$-foldings before reheating
(below we will set $N_{e}\approx60$), we find\begin{equation}
\phi_{1}=M_{\text{Pl}}\sqrt{\frac{nN_{e}}{2\pi}}\gg\phi_{*},\end{equation}
and then, using Eq.~(\ref{eq:PS def}) with $\phi=\phi_{1}$, we
get\begin{equation}
\lambda=\frac{3}{32\pi}n^{2}P_{S}\left(\frac{2\pi}{nN_{e}}\right)^{n+1}\ll1.\label{eq:lambda ans}\end{equation}

Substituting this value of $\lambda$ and Eqs.~(\ref{eq:H phi 2 d})--(\ref{eq:b phi 2 d})
into Eq.~(\ref{eq:b q equ}), we can obtain an implicit equation
for the value $\phi_{q}$. However, it is more convenient to express
$\phi_{q}$ through $b_{q}\equiv b(\phi_{q})$ using Eqs.~(\ref{eq:b phi 2 d})
and (\ref{eq:lambda ans}),\begin{equation}
\frac{M_{\text{Pl}}}{\phi_{q}}=P_{S}^{\frac{1}{2(n+1)}}\sqrt{\frac{2\pi}{nN_{e}}}b_{q}^{\frac{1}{n+1}},\label{eq:phi through b}\end{equation}
and to derive a closed-form equation for $b_{q}^{2}$,\begin{equation}
b_{q}^{2}=\frac{2}{n+1}\ln\frac{1}{P_{S}}+2\ln N_{e}-2\ln\left[\pi\left(n+1\right)b_{q}^{\frac{3n+5}{n+1}}\right].\label{eq:b q equ exact}\end{equation}
This equation is in a form that can be solved numerically by direct
iteration. We use the values $P_{S}=2.3\cdot10^{-9}$ and $N_{e}=60$,
while $n$ can be 1 or 2 \cite{Leach:2003us}, and obtain \begin{align*}
b_{q}^{2} & \approx13.9,\qquad n=1;\\
b_{q}^{2} & \approx8.9,\qquad n=2.\end{align*}
 A rough order-of-magnitude analytic expression for $b_{q}$ can be
obtained by using only the first term in Eq.~(\ref{eq:b q equ exact}),
\begin{equation}
b_{q}^{2}\approx\frac{2}{n+1}\ln\frac{1}{P_{S}}\approx\frac{40}{n+1}.\label{eq:bq rough}\end{equation}
However, this overestimates $b_{q}^{2}$ by about 50\%.

Let us now compute the average number of $e$-foldings during the
last NEC-preserving part of the trajectory, for potentials $V(\phi)\propto\phi^{2n}$.
According to the results of the previous section, we need to integrate
the $e$-foldings until the value $\phi_{q}$. Using Eq.~(\ref{eq:phi through b}),
we find\begin{align}
\left\langle \Delta N_{\text{NEC}}\right\rangle  & \approx\int_{\phi_{*}}^{\phi_{q}}d\phi\frac{H(\phi)}{-v(\phi)}=\frac{2\pi}{n}\frac{\phi_{q}^{2}}{M_{\text{Pl}}^{2}}\nonumber \\
 & \approx P_{S}^{-\frac{1}{n+1}}N_{e}b_{q}^{-\frac{2}{n+1}}\label{eq:DeltaN ans 01}\\
 & =O(1)P_{S}^{-\frac{1}{n+1}}N_{e}\left(\frac{2\ln P_{S}^{-1}}{n+1}\right)^{-\frac{1}{n+1}}.\label{eq:DeltaN ans 00}\end{align}
In the last line, we substituted for $b_{q}$ the rough estimate~(\ref{eq:bq rough})
merely in order to obtain a simpler analytic expression for $\left\langle \Delta N_{\text{NEC}}\right\rangle $
as an explicit function of the parameters. We use the more precise
Eq.~(\ref{eq:DeltaN ans 01}) for computing the numerical answers.
With the value $n=1$, which is preferred by observations, we find
\textbf{}\begin{equation}
\left\langle \Delta N_{\text{NEC}}\right\rangle \approx3.4\cdot10^{5}.\end{equation}
With $n=2$, we get $\left\langle \Delta N_{\text{NEC}}\right\rangle \approx2.2\cdot10^{4}$.

A curious coincidence is that Eq.~(\ref{eq:DeltaN ans 01}) can be
expressed through the slow-roll parameter $\varepsilon_{2}$ very
simply as\begin{equation}
\left\langle \Delta N_{\text{NEC}}\right\rangle \approx\frac{1}{\varepsilon_{2}(\phi_{q})}.\end{equation}
Here we need to compute $\varepsilon_{2}(\phi)$ at the value $\phi_{q}$
determined through $b_{q}$ as $b(\phi_{q})\equiv b_{q}$.

The value of $\phi_{q}$ that corresponds to the obtained value of
$b_{q}$ can be expressed as\begin{equation}
\frac{\phi_{q}}{M_{\text{Pl}}}=P_{S}^{-\frac{1}{n+1}}\frac{nN_{e}}{2\pi}b_{q}^{-\frac{2}{n+1}}=\frac{n}{2\pi}\left\langle \Delta N_{\text{NEC}}\right\rangle .\end{equation}
Thus, for $n=1$ we have $\phi_{q}\approx5.4\cdot10^{4}M_{\text{Pl}}$,
and for $n=2$ we have $\phi_{q}\approx7.0\cdot10^{3}M_{\text{Pl}}$.

We also need to check whether the intrinsic error of the present approximation,
as given by Eq.~(\ref{eq:delta Nq}), is small in comparison with
the mean value $\left\langle \Delta N_{\text{NEC}}\right\rangle $.
We get\begin{equation}
\delta N_{q}=\frac{\ln4}{\pi b_{q}^{2}\varepsilon_{2}(\phi_{q})}=\frac{\ln4}{\pi b_{q}^{2}}\left\langle \Delta N_{\text{NEC}}\right\rangle .\end{equation}
With the numerical values used above, we find $\delta N_{q}\sim10^{4}$
for $n=1$ and $\delta N_{q}\sim10^{3}$ for $n=2$. The relative
error of the approximation is given by $\ln4/(\pi b_{q}^{2})$ and
is about 3\% for $n=1$ and about 5\% for $n=2$, which is acceptable
for the purpose of our estimates.

\section*{Acknowledgments}

The author thanks Andrei Linde for many fruitful discussions.

\appendix

\section{Solving the stationary FP equation\label{sub:Solving-stationary-FP}}

The time integral of the propagator of the FP equation,\begin{equation}
\Psi(\phi_{\text{in}},\phi_{T})\equiv\int_{0}^{\infty}dT\, P(\phi_{\text{in}},\phi_{T},T),\end{equation}
can be computed in closed form in one-field models~\cite{Vilenkin:1999kd}.
One integrates Eq.~(\ref{eq:FP equ c1}) in time and finds\begin{align}
\hat{L}_{\phi}\int_{0}^{\infty}\negmedspace P(\phi_{0},\phi,t)dt & =\negmedspace\int_{0}^{\infty}\negmedspace\partial_{t}P(\phi_{0},\phi,t)dt\nonumber \\
 & =P(\phi,\infty)-P(\phi,0)=-\delta(\phi-\phi_{0})\end{align}
since $P(\phi,\infty)=0$. Hence, the function $\Psi(\phi_{0},\phi)$
satisfies the equation \begin{equation}
\hat{L}_{\phi}\Psi=-\delta(\phi-\phi_{0})\label{eq:psi L equ}\end{equation}
with the same boundary conditions in $\phi$ as the distribution $P(\phi_{0},\phi,t)$.
In other words, $\Psi$ is the Green's function of the stationary
FP equation. The function $\Psi$ can be computed by integrating Eq.~(\ref{eq:psi L equ}).
First, one obtains\begin{equation}
\partial_{\phi}\left(D\Psi\right)-v\Psi+\theta(\phi-\phi_{0})+C_{1}=0,\label{eq:psi 2}\end{equation}
where the integration constant $C_{1}$ is expressed using the boundary
condition~(\ref{eq:bc}) at $\phi=\phi_{\text{Pl}}$ (assuming $\phi_{0}>\phi_{*}$)
as\begin{equation}
C_{1}=-\theta(\phi_{\text{Pl}}-\phi_{0})=-1.\label{eq:c1 equ}\end{equation}
Finally, Eq.~(\ref{eq:psi 2}) can be integrated again; the general
solution can be written as \begin{align}
\Psi(\phi_{0},\phi) & =\frac{1}{D(\phi)\mu(\phi)}\left[C_{2}+\int_{\phi_{*}}^{\phi}\theta(\phi_{0}-\phi_{1})\mu(\phi_{1})d\phi_{1}\right],\label{eq:psi ans 0}\end{align}
where $C_{2}$ is an integration constant, and we introduced the auxiliary
function\begin{align}
\mu(\phi) & \equiv\exp\left[-\int_{\phi_{*}}^{\phi}\frac{v}{D}d\phi\right]\nonumber \\
 & =\exp\left[\frac{3M_{\text{Pl}}^{4}}{8}\left(\frac{1}{V(\phi_{*})}-\frac{1}{V(\phi)}\right)\right].\end{align}
The value of $C_{2}$ is determined through the boundary condition
at $\phi=\phi_{*}$. However, we note that $\mu(\phi)$ grows rapidly
with $\phi$; therefore the $C_{2}$ term is negligible for $\phi$
away from the reheating point. The closed-form expression for $\Psi$
is thus \begin{align}
\Psi(\phi_{0},\phi) & \approx\frac{1}{D(\phi)\mu(\phi)}\int_{\phi_{*}}^{\phi}\!\theta(\phi_{0}-\phi_{1})\mu(\phi_{1})d\phi_{1}\nonumber \\
 & =\frac{1}{D(\phi)}\negmedspace\int_{\phi_{*}}^{\min(\phi,\phi_{0})}\negthickspace d\phi_{1}\exp\left[-\negmedspace\int_{\phi}^{\phi_{1}}\negmedspace\frac{v(\phi_{2})}{D(\phi_{2})}d\phi_{2}\right].\label{eq:psi ans 1}\end{align}
We will now derive simplified forms of this expression in cases $\phi<\phi_{0}$
and $\phi>\phi_{0}$.

When $\phi<\phi_{0}$ (but for $\phi$ not too close to $\phi_{*}$),
we can simplify Eq.~(\ref{eq:psi ans 1}) if we note that the outer
integrand is dominated by the neighborhood of $\phi_{1}=\phi$ where
the exponent is close to 1. Then we can perform an asymptotic estimate
of the integral in Eq.~(\ref{eq:psi ans 1}). The easiest method
is to integrate by parts repeatedly, which yields an asymptotic series:
\begin{align}
\Psi(\phi_{0},\phi) & =\frac{1}{D(\phi)}\int_{\phi_{*}}^{\phi}\negthickspace d\left\{ -\frac{D(\phi_{1})}{v(\phi_{1})}\exp\left[-\int_{\phi}^{\phi_{1}}\frac{v}{D}d\phi_{2}\right]\right\} \nonumber \\
 & +\frac{1}{D(\phi)}\int_{\phi_{*}}^{\phi}\negthickspace\exp\left[-\int_{\phi}^{\phi_{1}}\frac{v}{D}d\phi_{2}\right]d\left\{ \frac{D(\phi_{1})}{v(\phi_{1})}\right\} \nonumber \\
 & \approx-\frac{1}{D(\phi)}\left\{ \frac{D(\phi)}{v(\phi)}+\frac{D(\phi)}{v(\phi)}\left(\frac{D(\phi)}{v(\phi)}\right)^{\prime}+...\right\} .\end{align}
Here we neglected the exponentially small terms of order\begin{equation}
\exp\left[-\int_{\phi_{*}}^{\phi}\left|\frac{v}{D}\right|d\phi_{2}\right]=\frac{1}{\mu(\phi)}\ll1.\end{equation}
Thus we find for $\phi<\phi_{0}$ the required result,\begin{equation}
\Psi(\phi_{0},\phi)\approx-\frac{1}{v(\phi)}\left[1+\left(\frac{D(\phi)}{v(\phi)}\right)^{\prime}+...\right]\approx-\frac{1}{v(\phi)}.\label{eq:psi approx ans}\end{equation}
In retaining only the first term of the asymptotic series, we neglect
terms involving $H/M_{\text{Pl}}\ll1$ as well as terms proportional
to the slow-roll parameters. 

For completeness, we give the result also for $\phi>\phi_{0}$. In
that case, the integral over $\phi_{1}$ in Eq.~(\ref{eq:psi ans 1})
becomes $\phi$-independent, and we get\begin{align}
\Psi(\phi_{0},\phi) & =\frac{\mu(\phi_{0})}{D(\phi)\mu(\phi)}\nonumber \\
 & =\frac{1}{D(\phi)}\exp\left[\frac{3M_{\text{Pl}}^{4}}{8}\left(\frac{1}{V(\phi_{0})}-\frac{1}{V(\phi)}\right)\right].\end{align}

\bibliographystyle{myphysrev}
\bibliography{EI2}

\end{document}